\documentclass[sigconf,natbib=true]{acmart} %

\usepackage{lipsum}
\AtBeginDocument{%
  \providecommand\BibTeX{{%
    \normalfont B\kern-0.5em{\scshape i\kern-0.25em b}\kern-0.8em\TeX}}}

\copyrightyear{2023}
\acmYear{2023}
\acmConference[Gen-IR@SIGIR2023]{The First Workshop on Generative Information Retrieval}{July 27, 2023}{Taipei, Taiwan}
\acmBooktitle{The First Workshop on Generative Information Retrieval (Gen-IR@SIGIR23), July 27, 2023, Taipei, Taiwan}
\acmDOI{}
\acmPrice{}
\acmISBN{}
\setcopyright{rightsretained}

\usepackage{enumitem}

\begin{document}

\title{Generative Retrieval as Dense Retrieval}

\author{Thong Nguyen}
\email{t.nguyen2@uva.nl}
\orcid{0000-0003-0607-0723}
\affiliation{%
  \institution{University of Amsterdam}
  \country{Netherlands}
}
\author{Andrew Yates}
\email{a.c.yates@uva.nl}
\orcid{0000-0002-5970-880X}
\affiliation{%
  \institution{University of Amsterdam}
  \country{Netherlands}
}

\begin{abstract}
Generative retrieval is a promising new neural retrieval paradigm that aims to optimize the retrieval pipeline by performing both indexing and retrieval with a single transformer model. However, this new paradigm faces challenges with updating the index and scaling to large collections.
In this paper, we analyze two prominent variants of generative retrieval and show that they can be conceptually viewed as bi-encoders for dense retrieval. Specifically, we analytically demonstrate that the generative retrieval process can be decomposed into dot products between query and document vectors, similar to dense retrieval. This analysis leads us to propose a new variant of generative retrieval, called Tied-Atomic, which addresses the updating and scaling issues by incorporating techniques from dense retrieval.
In experiments on two datasets, NQ320k and the full MSMARCO, we confirm that this approach does not reduce retrieval effectiveness while enabling the model to scale to large collections.

\end{abstract}

\maketitle

\section{Introduction}

Recent research has introduced several families of neural retrieval approaches that leverage transformer-based pre-trained language models, such as learned sparse retrieval, dense retrieval and cross-encoders \cite{lin2021pretrained}. The first two families encode queries and documents into sparse or dense vectors using transformer encoders, then use inverted or vector indexing for efficient retrieval. The cross-encoder, which is computationally expensive, is mainly used for re-ranking. 

A new paradigm known as generative retrieval or Differentiable Search Index (DSI) has recently emerged~\cite{tay2022transformer,wang2022neural}. This approach integrates indexing and retrieval processes within a single transformer stack, enabling joint optimization. During indexing, the model is trained to memorize the association between document texts and identifiers; during retrieval, it generates relevant document identifiers in response to a query. However, this approach has difficulty updating the index and does not scale well to large document collections~\cite{pradeep2023does,mehta2022dsi++}.
Most research has focused primarily on small collections with several hundred thousand documents.

Examining closely how generative retrieval models work, we can see that each generation step can be decomposed into the dot product of two vectors: the last hidden state emitted by the transformer's encoder-decoder stack and the vocabulary embeddings representing document identifiers. Conceptually, this process is analogous to dense retrieval where the relevance between a query and document is also modeled by the dot product between two vectors. This observation leads us to the key question in this paper: \textit{to what extent can generative retrieval be viewed as dense retrieval?}
\begin{figure}
    \centering
    \includegraphics[trim=200 195 150 170,clip,width=\linewidth]{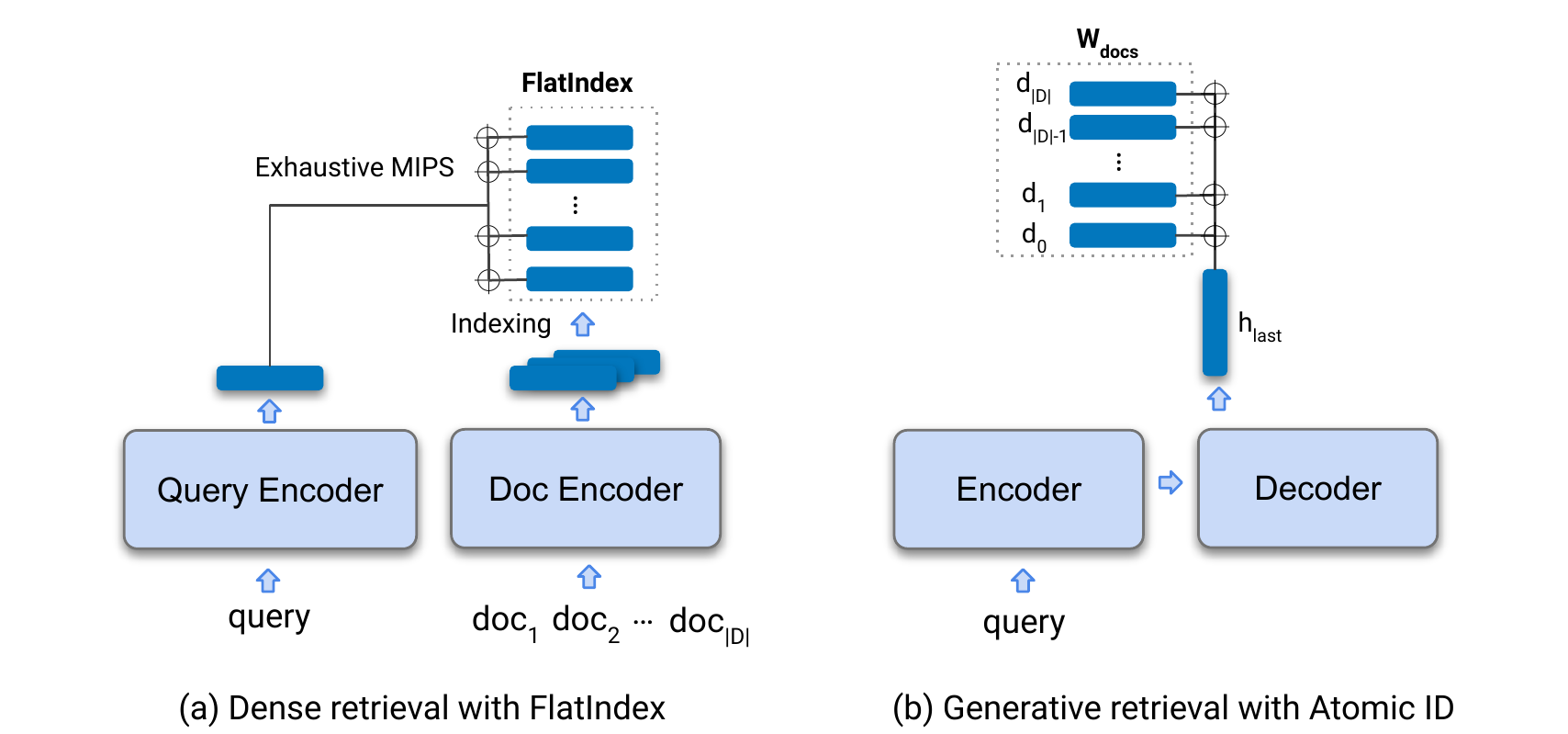}
    \caption{Comparing dense retrieval (using FlatIndex) with generative retrieval (using Atomic DocIDs).}
    \label{fig:dsi_atomic_dense}
    \vspace{-0.5cm}
\end{figure}
In this paper, we examine generative retrieval (DSI) models used with two document identifier/representation types: Atomic DocIDs and Hierarchical Semantic DocIDs \cite{tay2022transformer}. We show that these approaches are similar to dense retrieval in that they model relevance using the dot product between a query vector and a document vector.

Dense retrieval involves producing document vectors using a document encoder that takes in document text. Typically, the encoder is trained using a contrastive loss, which effectively guides the encoder to produce vectors that reflect the semantics of the documents. This learned text-vector association enables the encoder to generalize to new documents that were not seen during training. In contrast, generative retrieval treats document embeddings as randomly initialized model parameters and tries to align these vectors with corresponding texts using teacher forcing techniques during training.
Empirically, generative retrieval approaches face challenges when generalizing to unseen documents or scaling up to large collections~\cite{pradeep2023does,mehta2022dsi++}. Adding new documents requires the model parameters to be optimized again, and scaling up to larger collections may require increasing the model size.

To overcome these limitations, we propose the Tied-Atomic model, which enhances the DSI generative retrieval model with Atomic DocIDs by (1) tying document texts to vocabulary (DocID) embeddings and (2) using a contrastive loss for training. Our experiments demonstrate that our proposed model performs competitively with the DSI model and NCI model~\cite{wang2022neural} with Hierarchical Semantic DocIDs on the NQ320K dataset. Additionally, our model can handle unseen documents\footnote{This trivially comes ``for free'' as a consequence of the dense retrieval view.} and scaled up for large collections, as we demonstrate with the full MSMARCO dataset.




\section{Related Work}
\textbf{Dense retrieval}~\cite{karpukhin2020dense, DBLP:conf/iclr/XiongXLTLBAO21, DBLP:conf/sigir/HofstatterLYLH21} is a class of first-stage retrieval methods that use bi-encoders to encode queries/documents into dense vectors and compute the relevance between a query and a document as the dot product between its dense vectors. Given a query vector, a dense retrieval system computes the dot product between the query vector and candidate document vectors, then returns the top-k documents with the highest scores. To speed up the retrieval process, dense retrieval is usually accompanied with a vector index. Thanks to the separation of the query and document encoding process, document vectors can be computed and indexed offline. There are different choices of index~\cite{cvpr20_tutorial_image_retrieval} (e.g., Tree-based Index~\cite{ram2013space, zhu2019joint, liu2004investigation}, Graph-based Index~\cite{DBLP:journals/pami/MalkovY20}), but the main idea of these indexes is to step-by-step narrow down the search space or the number of documents to be processed. Using an index usually leads to a reduction in effectiveness in exchange for efficiency. 



\textbf{Generative retrieval}, also known as Differentiable Search Index, is an emerging retrieval paradigm that aims to generate relevant document identifiers (DocID) directly without relying on a separate index~\cite{metzler21rethinking,tay2022transformer}. One of the pioneers of this approach is \citet{tay2022transformer}, who proposed using generative transformers like the T5 transformer~\cite{DBLP:journals/jmlr/RaffelSRLNMZLL20} for both indexing and retrieving.

During the indexing phase, the model is trained to memorize the document collection and align each document text with its corresponding DocID. At the fine-tuning step, the model generates relevant DocIDs from query text. Both indexing and fine-tuning use the seq2seq cross-entropy loss with teacher forcing~\cite{DBLP:journals/neco/WilliamsZ89} to update the model's parameters. \citet{tay2022transformer} reported that training indexing and fine-tuning together in a multi-task setup typically yields better results than training the tasks separately. To further improve the generative model's performance, recent studies ~\cite{wang2022neural, zhou2022ultron, sun2023learning} have used generated queries from docT5query as data augmentation. docT5query~\cite{Nogueira2019DocumentEB} is a method for generating synthesized queries from a single input query using the T5 transformer~\cite{DBLP:journals/jmlr/RaffelSRLNMZLL20}. \citet{zhuang2022bridging} also showed that query generation helps bridging the query distribution gap between indexing and retrieval phrase.

During the retrieval phase, the beam search algorithm is used to decode the most likely DocIDs. The generation probability of each DocID is then used for ranking. The efficiency of this decoding step is greatly influenced by how document identifiers are structured. Different approaches have been proposed for representing document identifiers, including Atomic DocIDs and Semantic Structured DocIDs, also known as Hierarchical Semantic DocIDs \cite{wang2022neural}. 
With Atomic DocIDs, each document is assigned an arbitrary unique token that extends the transformer's vocabulary. On the other hand, Semantic Structured DocIDs use hierarchical clustering algorithms, such as k-means, to group semantically similar documents into a tree/trie. Each document is identified by a sequence of nodes from the root to the corresponding leaf node, where nodes are represented by a set of extended vocabulary in the transformer~\cite{DBLP:journals/jmlr/RaffelSRLNMZLL20}.

\section{Generative vs. dense retrieval}
The main distinguishing factor between generative retrieval variants is how they construct document identifiers (DocIDs). This section will discuss prominent document identifier variations and discuss their similarities to the dense retrieval paradigm.

\subsection{Atomic DocIDs}
Unstructured Atomic Identifiers (Atomic DocIDs) assign each document a unique identifier represented by a new token added to the transformer's vocabulary. Initially, the embeddings of new tokens are randomly initialized, but during indexing and fine-tuning, they are updated to better reflect document content. During inference, the model performs a single decoding step that produces a vector of generation logits or probabilities over the new vocabulary tokens. The relevance of a document to a query is the generation probability of the corresponding document identifier. Formally, the output probability over documents is generated as follows:
\begin{align}
O = \text{Softmax}([W_{\text{docs}}]^T h_{\text{last}})
\end{align}

In this equation, $W_{\text{docs}}$ is the document embedding matrix, where each row represents the embedding of a document identifier, and $h_{\text{last}}$ is the last hidden state of the transformer's decoder stack. The Softmax function can be removed without affecting the document ranking. As we can see from the equation, the score between a query and a document is computed as the dot product between two vectors: the hidden state $h_{\text{last}}$ and the token embedding (rows in $W_{docs}$). This suggests that generative retrieval with Atomic DocIDs can closely correspond to dense retrieval, as shown in Figure \ref{fig:dsi_atomic_dense}. The last hidden state $h_{\text{last}}$ and the token embedding can be considered the query and document vectors, respectively, in dense retrieval.
It follows that generative retrieval with Atomic DocIDs is no more expressive than dense retrieval with bi-encoders. In both cases, the query-document similarity computation reduces to a dot product between two independent vectors.

Approaches for generative retrieval with Atomic DocIDs and dense retrieval do exhibit several differences, such as in how document vectors are created. In generative retrieval, document vectors are initialized as separate vocabulary embeddings and optimized via the indexing and training steps. This is different from dense retrieval, where the document vectors are the output of a document encoder that receives document texts as input. The dense retrieval encoder ties document texts and vectors together in a sense, allowing it to generalize to similar and unseen documents. In contrast, atomic generative retrieval separates document texts from token embeddings, leaving the model with no way to create meaningful embeddings for unseen documents.


Dense retrieval and atomic generative retrieval typically use different training objectives. Dense retrieval uses a supervised contrastive loss, which requires positive and negative documents per training query. It pulls the positive document's vectors closer to the query vectors and pushes negative documents further away. A contrastive loss with hard negatives has shown to be very important for training an effective dense retrieval model. On the other hand, atomic generative retrieval uses teacher forcing with a seq-2-seq cross entropy loss, which increases the likelihood of a chosen positive documents (and thus pushes away all other documents).

\subsection{Hierarchical Semantic DocIDs}
\begin{figure}[h!]
    \centering
    \includegraphics[trim=150 225 460 210,clip,width=0.6\linewidth]{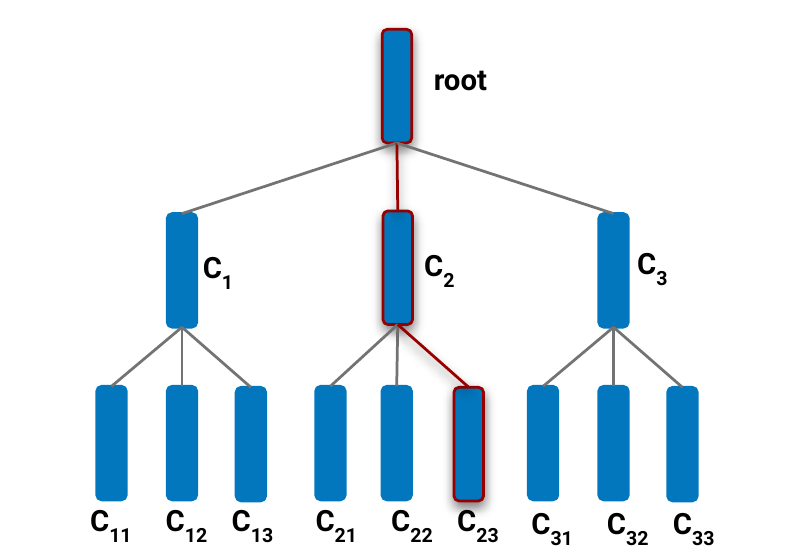}
    \caption{Hierarchical semantic identifiers/ tree index.}
    \label{fig:tree_index}
\end{figure}

To leverage similarities among documents and address the inefficiency of Atomic DocIDs, researchers~\cite{tay2022transformer,wang2022neural} proposed Hierarchical Semantic DocIDs that could (1) capture document semantics through document clusters and (2) improve efficiency by narrowing down the search space after each decoding step. 
To construct Hierarchical Semantic DocIDs, the document collection is recursively clustered (using document embeddings) into $C$ clusters until each cluster has no more than $C$ items, resulting in a tree structure as in Figure \ref{fig:tree_index}. At each level, nodes from the same parent are labeled with numbers from \textit{1} to \textit{C}. In the tree, each document is identified by the path from the root to the corresponding leaf node; this path is represented by a sequence of numbers where the $i^{th}$ number denotes the node at $i^{th}$ level. 

\begin{figure}[h!]
    \centering
    \includegraphics[trim=400 190 110 185,clip,width=0.8\linewidth]{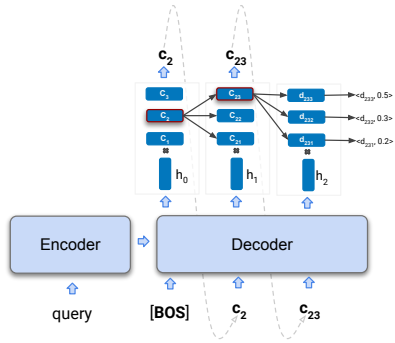}
    \caption{Generative retrieval with hierarchical semantic docids as a form of dense retrieval.}
    \label{fig:decoding}
\end{figure}

To support identifier generation, $C$ new tokens, each representing a cluster label, are added into the generative model. The model performs multiple decoding steps to generate document Hierarchical Semantic DocIDs as in Figure \ref{fig:decoding}. At the first step, the decoder stack produces the first hidden state $h_0$ which is then used to select the most likely first-level cluster (using MIPS). The selected first-level cluster is then fed back to the decoder input to generate the second hidden state and pick the next-level cluster. This process continues until a leaf node is reached and the most likely DocID is then returned. After each decoding step, the search space is reduced to one branch of the tree; the hidden state and token embeddings can also be refined. Depending on the configuration of the decoding algorithm (e.g., beam search), the number of clusters selected at each decoding step vary. 

From a broad perspective, this variant of generative retrieval is similar to dense retrieval with a tree-index. The tree index is implemented in popular ANN libraries like FAISS; it recursively split the search space into smaller subspaces, forming a trie as in Figure \ref{fig:tree_index}. During retrieval, only the $n$ closest subspaces (based on the distance between the query vector and subspace's centroid) at each tree level are explored. For comparison, the hidden states and token embeddings in semantic generative retrieval can be thought of as the query and centroid vectors in the tree index's search. The differentiating factor of this class of generative models could be their ability to refine or adapt query embeddings (hidden states) or centroid (token) embeddings as the search progresses further down the tree. However, the impact of embedding refinement on retrieval effectiveness remains uncertain and requires further investigation.
While this document identifier/representation is theoretically more expressive than Atomic DocIDs, current approaches using Hierarchical Semantic DocIDs do not outperform Atomic DocIDs~\cite{pradeep2023does}.

\section{Methodology: Tied-Atomic}
\begin{figure}[h!]
    \centering
    \includegraphics[trim=300 220 280 180,clip,width=0.6\linewidth]{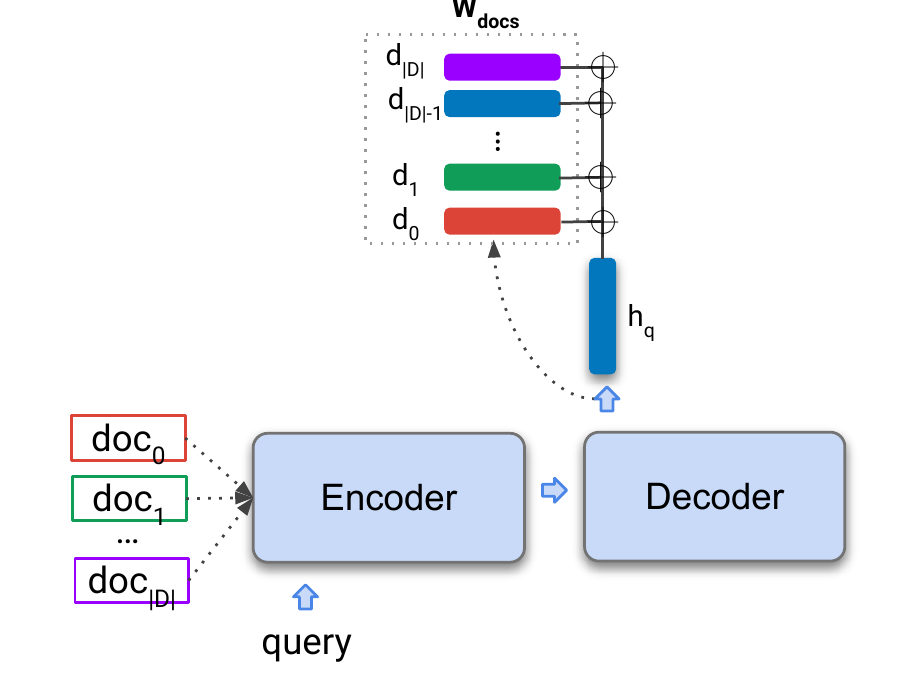}
    \caption{Tied-Atomic: token (DocID) embeddings are tied to the document texts via the encoder-decoder stack.}
    \label{fig:tied-atomic}
\end{figure}

Based on the analysis in the previous section, generative retrieval can be seen as a form of dense retrieval, with a new adaptive query processing algorithm in the case of Hierarchical Semantic DocIDs. Viewing generative retrieval from the perspective of dense retrieval opens opportunities to incorporate techniques from dense retrieval to address the issues (e.g., with updating and scaling) of generative retrieval. To demonstrate this, we proposes the following modifications to generative retrieval (DSI) with Atomic DocIDs:

\begin{itemize}[leftmargin=*]
    \item We tie the token (DocID) embeddings in the T5 vocabulary to the corresponding input document texts. This forces DocID embeddings to be the output of T5's encoder-decoder stack rather than the model's parameters. Given this tying, the document ranking scores are computed as followings: 
    \begin{align}
O = \text{Softmax}([W_{\text{docs}}]^T h_{\text{query}})
\end{align}
where $W_{\text{docs}} = [h_{\text{doc1}}, h_{\text{doc2}}, \text{...},h_{\text{docD}}]^T$ and $h_{\text{query}}$/$h_{\text{doc\{x\}}}$ are the output vectors (last hidden states) of T5's encoder-decoder stack with corresponding query and document texts as inputs. This change is also illustrated in Figure \ref{fig:tied-atomic} comparing to the original model shown in Figure \ref{fig:dsi_atomic_dense}.
This text embedding approach is analogous to SentenceT5's EncDec variant~\cite{ni-etal-2022-sentence}.

\item In the original DSI setup, the computation of training loss requires a logit vector over the entire decoder's vocabulary. This process is particularly expensive for Atomic DocIDs where the decoder's vocabulary is the entire document collection. To reduce the computation, we employ the contrastive loss with BM25 negatives (including in-batch negatives) sampled from top-k ranked documents, which is typically used in dense retrieval training.
\end{itemize}

These modifications only affect the training phase; the computation inside the model remains the same at the retrieval step with embeddings untied. With these changes, atomic generative models can theoretically insert new documents or scale to large collections as easily as dense retrieval methods.
While inference is performed in a generative way, this model is a bi-encoder in some sense.
However, with these modifications the model still has the same expressiveness as generative retrieval using Atomic DocIDs!


\section{Experiments and Results}
\subsection{Datasets}
We conduct experiments on two main datasets: NQ320K~\cite{kwiatkowski2019natural} and MSMARCO passages~\cite{DBLP:conf/nips/NguyenRSGTMD16}. The NQ320K dataset has approximately 307k query-document training pairs and approximately 8k development pairs. We used the scripts provided by \citet{wang2022neural} to preprocess the document collections, which includes removing HTML tags from the documents and deduplicating documents using titles. The resulting document collection contains more than 100k relevant documents gathered from all training and development pairs. MSMARCO is a much larger and more popular dataset for training and evaluating information retrieval methods. It contains more than 8 million short passages, approximately 809k training queries, and approximately 7k development queries. Most of the training and development queries have one relevant document.

\subsection{Experimental setup}
On the NQ320K dataset, we trained our Tied-Atomic T5-base model for 120k steps (batch\_size=32) using a ranking loss with in-batch BM25 negatives similar to ~\cite{formal2021splade}. Additionally, we trained a dense model with a distillBERT\cite{sanh2019distilbert} encoder using similar configurations for an additional comparison point. An optimized version of BM25 and BM25+docT5query was provided; we optimized the values of $k1$ and $b$ on a small subset of training data using the Anserini toolkit~\cite{yang2017anserini}. For the MSMARCO dataset, we trained our Tied-Atomic model using the Sentence-Transformer's mined hard negatives and CE scores for distillation training with MarginMSELoss~\cite{reimers-2019-sentence-bert, DBLP:journals/corr/abs-2010-02666}.
We freeze Tied-Atomic DocIDs during training. While one could copy the output document vectors back to the model's (vocabulary) parameters and continue fine-tuning the model, we expect doing so would hinder the model's ability to generalize to unseen documents.

\subsection{Results and discussion}
Table \ref{tab:nq320k} presents the results of our Tied-Atomic approach, the DSI and NCI generative retrieval models, a dense retrieval bi-encoder, and lexical baselines on the NQ320k dataset. Variations of this dataset have been commonly used in generative retrieval studies due to its small size~\cite{tay2022transformer, wang2022neural, zhou2022ultron}. 

On this dataset, we can see that BM25 with optimized parameters ($k1=10, b=0.8$) already achieves a high R@1 score of 45.95, meaning that nearly half of the queries could be solved with only lexical matching.
Expanding documents with queries generated by the docT5query model improves BM25 scores consistently on all metrics. The NCI (base) model, which also uses Hierarchical Semantic DocIDs together with adaptive vocabulary embeddings, performs significantly better than the BM25 and DSI baselines in most metrics. NCI (base) is only comparable to BM25 and slightly worse than docT5query at R@100. 

Our Tied-Atomic (base) performs slightly worse than NCI (base) at R@1 but slightly better on other metrics, with improvements ranging from around 2\% to around 6\%. Note that our Tied-Atomic uses Atomic DocIDs, which were reported to perform much worse than Hierarchical Semantic DocIDs \cite{tay2022transformer} used in the NCI (base) model. We attribute the improvement of Tied-Atomic to the changes (embedding-ties, contrastive loss) that make training more efficient. More interestingly, we find that both Tied-Atomic using atomic identifiers and NCI (base) using hierarchical semantic identifiers perform similarly to dense retrieval with a DistilBERT encoder using 50\% fewer parameters than T5-base. This result is in contrast to findings recently reported by \citet{wang2022neural} and \citet{tay2022transformer}, who obtain much worse performance using dense retrieval.

\begin{table}[h!]
    \centering
    \caption{Results on the NQ320k collection.}
    \resizebox{\linewidth}{!}{\begin{tabular}{lcccc}
    \toprule
    \toprule
    \textbf{Method} & \textbf{R@1} & \textbf{R@10} & \textbf{R@100} & \textbf{MRR@100} \\ 
    \midrule
    BM25 & 45.95 & 78.19 & 92.69 & 57.23 \\ 
    BM25  + docT5Query & 48.74 & 80.29 & 94.05 & 59.88 \\ 
    DSI Semantic (base)~\cite{pradeep2023does} & 58.70 & - & - & - \\ 
    DSI Atomic (base)~\cite{pradeep2023does} & 60.00 & - & - & - \\ 
    NCI (base)~\cite{wang2022neural} & \textbf{65.86} & 85.20 & 92.42 & 73.21 \\ 
    \midrule
    Dense (distilbert) & 64.46 & 88.13 & 95.18 & 73.33  \\ 
    \midrule
    Tied-Atomic (base) & 65.26 & \textbf{90.03} & \textbf{96.16} & \textbf{74.67} \\ 
    \bottomrule
    \end{tabular}}
    \label{tab:nq320k}
\end{table}

The results on NQ320k suggest that we do not lose any effectiveness by viewing generative retrieval from the perspective of dense retrieval.
To demonstrate that this approach improves scalability, we trained our Tied-Atomic model initialized from T5-small on the MSMARCO collection with $\sim$8 million documents. 
As shown in Table \ref{tab:msmarco}, Tied-Atomic performs reasonably on MSMARCO with comparable performance to ANCE~\cite{DBLP:conf/iclr/XiongXLTLBAO21}, a well-performing dense retrieval model. Despite using a small T5 model, Tied-Atomic outperforms DSI Atomic (base) \cite{pradeep2023does} by 38\% in term of MRR@10.

\begin{table}[ht!]
    \caption{Results on the full MSMARCO collection.}
     \centering
     \begin{tabular}{lcc}
     \toprule
     \toprule
     \textbf{Method}  & \textbf{MRR@10} & \textbf{R@1000} \\
     \toprule
        BM25 & 18.92 &  85.55 \\
        BM25 + docT5query & 27.70 & 94.70 \\ 
        ANCE \cite{DBLP:conf/iclr/XiongXLTLBAO21}& 33.00 & 95.90 \\
        DSI Atomic (base)~\cite{pradeep2023does}& 24.20 & - \\ 
        \midrule
        Tied-Atomic (small) & \textbf{33.42} & \textbf{96.47} \\ 
    \bottomrule
     \end{tabular}
     \label{tab:msmarco}
     \vspace{-0.4cm}
 \end{table}
\section{Conclusion}
In this paper, we analyze the new generative retrieval paradigm and show how it can be viewed as a form of dense retrieval.  The insights from this analysis led us to propose a generative retrieval model with a new document representation/identifier, Tied-Atomic, that performs competitively with previous generative models while also allowing for updating of the index and scalability to large document collections. While our experiments mainly focus on atomic identifiers, experiments in recent work suggest that other existing identifiers do not outperform atomic identifiers \cite{pradeep2023does}.


\bibliographystyle{ACM-Reference-Format}
\bibliography{sample-base}

\end{document}